\documentclass[pra,amsmath,amsfonts,superscriptaddress,showpacs,twocolumn]{revtex4-1}

\usepackage[bookmarksnumbered, colorlinks, plainpages]{hyperref}
\usepackage{epsf,epsfig}
\usepackage[psamsfonts]{amssymb}
\usepackage{amsmath}
\usepackage{graphicx}
\usepackage{color,soul}
\usepackage{xcolor}
\usepackage{pgf,tikz}
\usepackage{graphicx}
\usepackage{dcolumn}
\usepackage{bm}
\usepackage{color,soul}
\usepackage{xcolor}	
\usepackage[hang,raggedright]{subfigure}
\usepackage{csquotes}

\usepackage{bbold}
\graphicspath{{./fig/}}

\newtheorem{theorem}{Theorem}
\newtheorem{lemma}{Lemma}

\newtheorem{observation}[theorem]{Observation}

 \newcommand{\ket}[1]{|#1\rangle}
 \newcommand{\bra}[1]{\langle #1|}
 \newcommand{\project}[1]{\ket{#1}\bra{#1}}

 \newcommand{\Id}{{\mathbb 1}}
 \newcommand{\e}{{\mathrm e}}

 \newcommand{\T}{{\mathrm t}}

\newcommand{\dif}{{\mathrm d}}
\newcommand{\Tr}{{\mathrm {Tr}}}
\newcommand{\diag}{{\mathrm {diag}}}

 \newcommand*{\rom}[1]{\expandafter\@slowromancap\romannumeral #1@}

\begin{document}
\title{Quantumness  of quantum channels }
\author{Fereshte Shahbeigi}
\email{fereshte.shahbeigi@mail.um.ac.ir}
\author{ Seyed Javad Akhtarshenas}
\email{akhtarshenas@um.ac.ir}

\affiliation{Department of Physics, Ferdowsi University of Mashhad, Mashhad, Iran}

\begin{abstract}
Quantum coherence is a fundamental aspect of quantum physics  and plays a central role in  quantum information science.
This essential property  of the quantum states could be  fragile under the influence of the quantum operations. The extent to which quantum coherence is diminished depends both on the channel and the incoherent  basis.  Motivated by this, we propose a measure of nonclassicality of a quantum channel  as the average quantum coherence of the state space after the channel acts on, minimized over all orthonormal basis sets   of the state space. Utilizing the squared $l_1$-norm of coherence for the qubit channels, the minimization can be treated analytically and the proposed measure takes a closed form of expression. If we allow the channels to act locally on a  maximally entangled state,  the quantum correlation is diminished   making the states  more classical. We show that the extent to which quantum correlation is preserved under local  action of the channel cannot exceed  the quantumness of the underlying  channel. We further apply  our measure to the quantum teleportation protocol and show that a nonzero quantumness for the underlying  channel provides a necessary condition to overcome the best classical protocols.

\end{abstract}

\maketitle


\section{Introduction}\label{introduction}
 Quantum coherence arising from superposition is the most fundamental aspect of quantum mechanics and is responsible for the nonclassical properties of quantum systems. The role of quantum coherence in the wide range of areas, e.g. thermodynamics, quantum algorithms, quantum correlations, biology, etc. \cite{streltsovrev2017,streltsovprl2015,streltsov distil,napoli2016,anand,hillery2016,shi2017,piani2016,bu2016,karpat2014,marvian2016,marvian2016 2} shows the importance of its investigation.   Although it is not a new concept and is as long-standing as quantum mechanics itself, the study of quantum coherence as a quantum resource theory (QRT) \cite{pleniocoherence2014} has attracted a lot of interest recently.

On the other hand it is understood that  quantum channels, the completely positive and trace preserving maps, could reduce  quantum coherence of the states.  Quantum channels provide a framework to investigate coherence behavior under different dynamics and noises.  Cohering power and decohering power  \cite{manikarimipour,coheringpower}  in analogy to entangling power \cite{entangling1,entangling2} for quantum operations and also  classification of quantum channels based on the coherence-breaking property \cite{coherencebreaking} in comparison with entanglement-breaking channels \cite{entanglementbreaking} arose from these investigations.
Another important class of channels is the so-called semiclassical channels,  those that map all input quantum states into ones which are diagonal in the same basis; $\forall\rho \;:\; \mathcal{E}(\rho)=\sum_n p_n(\rho)\project n$ \cite{streltsovprl2011}.  Clearly, as a result of applying  a semiclassical channel the space of quantum states changes into a classical one,  meaning  that  all states are distinguishable by a classical observer measuring states with respect to the channel's diagonal basis \cite{meznaric 2013}.

The QRT provides preliminaries to recognize the interrelations between coherence and different  physical theories  \cite{streltsovrev2017,streltsovprl2015,streltsov distil,chitambar coh vs ent,zhu 2017,streltsovprx2017,streltsov merging,coh vs cor, frozencoh}. Particularly, it has been shown that  coherence is the resource in the process of generating entanglement  in bipartite systems \cite{streltsovrev2017,streltsovprl2015}.  
Furthermore, one of the most significant results is that quantum discord type correlations are simply the minimum amount of quantum coherence of the bipartite system  with respect to the  local  product basis \cite{frozencoh}.  In view of this, the inability of semiclassical channels  in the local creation of quantum correlations is predictable  as they remove all coherence of the state they are applied on. Actually,  it has been shown  that for the creation of quantum correlation via local channels in finite-dimensional systems, a necessary and sufficient condition for channels is that they should not be commutativity preserving  \cite{hu 2012}. A quantum channel $\mathcal{E}$ is said to be a commutativity preserving channel if it preserves the commutativity of any two compatible states, i.e., $\forall\;\rho,\xi  :  [\rho,\xi]=0$, then $[\mathcal{E}(\rho),\mathcal{E}(\xi)]=0$ \cite{hu 2012}. Clearly semiclassical channels are commutativity preserving, so nonsemiclassicality   is a necessary condition to create quantum correlations locally. Particularly in qubit systems, the necessary and sufficient conditions for channels are nonunitality and nonsemiclassicality \cite{streltsovprl2011}. These results lead us to conclude that there should be connections between the amount of nonsemiclassicality of one-party channels and quantum correlations of the bipartite states.

Furthermore, one can claim that  nonunitary quantum channels generally induce decay in quantum properties of the states with respect to some  orthonormal basis  that the coherence of most outputs is reduced.   This is at the heart  of the channels' contraction effect on the space of states, the case of semiclassical channels is the special one making the states completely classical. In other words,  a semiclassical channel washes  out quantum coherence of  the input states  making  the  states  distinguishable by a classical observer.
We adapt this as  our notion of   classicality of channels and pursue the question of  how far  a channel is from this very notion of classicality.
 Our notion of classicality is different from the one given in \cite{meznaric 2013} based on the concept of einselection, characterizing  a classical operation as a one that commutes with the completely dephasing process. Here we introduce a measure to quantify the amount of quantumness preserved by the channel, along with a natural coherence-based  criterion to decide whether or not a given channel is  semiclassical. We further investigate the local action of the channels and show that the quantumness of the channel provides an upper bound for the quantum correlations of the associated Choi states. Moreover, by applying the measure to quantum teleportation protocol, we show that the quantumness of the underlying channel provides a necessary, but not sufficient, condition to overcome the best classical protocols.

The paper is organized as follows. In Sec. \ref{nonsemiclassicality}, we introduce our method to quantify the quantumness of the channels. A closed form for the quantumness of the qubit channels and its  properties are also presented in Sec. \ref{nonsemiclassicality}. Section \ref{correlations} is devoted to  study local action of the qubit channels. In particular, we identify a  relation between the channel's  quantumness and the quantum correlation  of its associated Choi state. Application of the introduced  measure to quantum teleportation protocol is presented  in Sec.  \ref{Section-Application}.  The paper is concluded in Sec.  \ref{conclusion} with a brief discussion.

\section{Quantumness of the quantum channels }\label{nonsemiclassicality}
Quantum coherence is  evidence of the existence of the quantum properties in a given state.  In order to investigate the amount of quantumness of the state space,  a good candidate is to  average on coherence of all states. Such an average indicates the quantumness of the state space with respect to an assumed incoherent basis. It turns out that  minimization  over all orthonormal basis sets results in  finding the one  that most of the states behave more classically with respect to it. Obviously, without any evolution in states, the state space is so symmetric that the minimization is completely pointless. The importance of such optimization appears in the case of the nonunitary evolution of states and the result addresses the question of  how much quantumness  a quantum channel preserves. Clearly, this is a restatement of asking  how much a quantum channel deviates from a semiclassical one, as this class of operations erases the quantum coherence of all states completely. Motivated by this, we define the  quantumness (or, equivalently, the nonsemiclassicality)  of a quantum channel as
\begin{equation}\label{NS}
\mathcal{Q}_{C}(\mathcal{E})=N_C\min_{\{\ket i\}}\int C\big(\mathcal{E}(\rho)\big)d\mu(\rho),
\end{equation}
where subindex $C$ denotes that the defined quantity depends on the chosen measure of coherence,  and  the factor $N_C$ is introduced in order to normalize the desired quantity at the end.
Above, the integration is to average on coherence of all states.  Moreover, regarding that coherence is a base dependent concept, the minimization taken over all of the orthonormal basis ensures that $\mathcal{Q}_{C}(\mathcal{E})$ is independent of the particular choice of the incoherent basis. If we skip the optimization of Eq. \eqref{NS}, the resultant quantity represents the deviation of the quantum channel $\mathcal{E}$ from coherence breaking maps \cite{coherencebreaking} with respect to an assumed incoherent basis. Such a quantity, intuitively,  cannot be less than $\mathcal{Q}_{C}(\mathcal{E})$ for a general quantum channel $\mathcal{E}$.      It could be straightforwardly proved, because of the non-negativity condition of any bona fide measure of coherence, that $\mathcal{Q}_{C}(\mathcal{E})\geq0$ and it is zero if and only if the quantum channel is semiclassical. Furthermore, as a result of integration over all input states, the equality $\mathcal{Q}_{C}(\mathcal{E}\circ\mathcal{U})=\mathcal{Q}_{C}(\mathcal{E})$ holds and minimization over  all incoherent bases leads to $\mathcal{Q}_{C}(\mathcal{U}\circ\mathcal{E})=\mathcal{Q}_{C}(\mathcal{E})$, where $\mathcal{U}$ is an arbitrary unitary map, $\mathcal{U}(\cdot)=U\cdot U^\dagger$ for unitary  operator $U$. So as an immediate  corollary, the nonsemiclassicality of any unitary map is the same as the one of identity channel $\mathfrak{I}$ with no action on its input states, $\mathcal{Q}_{C}(\mathcal{U})=\mathcal{Q}_{C}(\mathcal{U}\circ\mathfrak{I})=\mathcal{Q}_{C}(\mathfrak{I}\circ\mathcal{U})=\mathcal{Q}_{C}(\mathfrak{I})
$.

Before we  proceed further with Eq. \eqref{NS}, we have to choose a suitable measure of coherence \cite{napoli2016,streltsovrev2017,streltsovprl2015,winter2016,zhu 2017,yu 2016,shao 2015,pleniocoherence2014,Aberg2006}. One of  such measures  that satisfies all conditions imposed by QRT  is the so called $l_1$-norm of coherence \cite{pleniocoherence2014} defined by $C_{l_1}(\rho)=\sum_{i,j \atop{i\ne j}}|\rho_{ij}|$. On the other hand, it is shown that   the sum of the squared absolute values of all off-diagonal elements violates monotonicity \cite{pleniocoherence2014}, however for qubit systems  it coincides, up to  a normalization factor $1/2$, with the squared $l_1$-norm of coherence,
$C^2_{l_1}(\rho)$,
and satisfies the properties of a bona fide measure of coherence (see Appendix \ref{Appendix1} for the proof).
In the following section we focus on qubit channels and show that the minimization specified in Eq. \eqref{NS} can be calculated analytically for the squared $l_1$-norm of coherence.

\subsection{Qubit channels}
The effect of the qubit channel $\mathcal{E}$ is an affine transformation on the Bloch sphere; $\mathcal{E}: \boldsymbol{r}\longrightarrow \boldsymbol{r}^\prime= \Lambda\boldsymbol{r}+\boldsymbol{t}$, where  $\boldsymbol{r}$ is the Bloch vector of the input state. Here  the  $3\times3$ matrix $\Lambda$   and the  vector $\boldsymbol{t}$  are  features of the  channel $\mathcal{E}$.  Having this in mind, to be more computable, we  choose the squared  $l_1$-norm   as the measure of coherence and derive   an explicit expression for the quantumness of  qubit channels.
To apply Eq. \eqref{NS} in this case, one has to measure the quantum coherence of any  state $\mathcal{E}(\rho)=(\Id_2+\boldsymbol{r}^\prime \cdot\boldsymbol{\sigma})/2$ with respect to an arbitrary orthonormal  basis $\{\ket{n+},\ket{n-}\}$, where $\ket{n+}=\cos\frac{\alpha}{2}\ket0+\e^{i\beta}\sin\frac{\alpha}{2}\ket1$ and $\ket{n-}=\e^{-i\beta}\sin\frac{\alpha}{2}\ket0-\cos\frac{\alpha}{2}\ket1$. A straightforward calculation shows that
\begin{eqnarray}\nonumber
 C_{l_1}^2\big(\mathcal{E}(\rho)\big)&=&4|\bra{n-}\mathcal{E}(\rho)\ket{n+}|^2=|\boldsymbol{r}^\prime\times\hat{\boldsymbol{n}}|^2 \\  \label{outputcoherence} &=&|(\Lambda\boldsymbol{r}+\boldsymbol{t})\times\hat{\boldsymbol{n}}|^2,
\end{eqnarray}
which can be used to obtain the  quantumness  of the   qubit channels
\begin{eqnarray}\nonumber
\mathcal{Q}_{C_{l_1}^2}(\mathcal{E})&=&\frac{5}{2}\frac{3}{4\pi}\min_{\hat{\boldsymbol{n}}}\int |(\Lambda\boldsymbol{r}+\boldsymbol{t})\times\hat{\boldsymbol{n}}|^2\dif^3r \\ \label{NSqubit}
&=& \frac{5}{2}\frac{3}{4\pi}\min_{\hat{\boldsymbol{n}}}
\int\big(|\Lambda\boldsymbol{r}\times\hat{\boldsymbol{n}}|^2+|\boldsymbol{t}\times\hat{\boldsymbol{n}}|^2\big)\dif^3r.
\end{eqnarray}
Above  we set $N_{C_{l_1}^2}=5/2$  and the minimum is taken over all unit vectors $\hat{\boldsymbol{n}}\in \mathbb{R}^3$. Also the last equality follows  from $\int(\Lambda\boldsymbol{r}\times\hat{\boldsymbol{n}})\cdot(\boldsymbol{t}\times\hat{\boldsymbol{n}})d^3r=0$, arisen  from  the fact that  for any  $\boldsymbol{r}$ in the integration  there is a corresponding  $-\boldsymbol{r}$.  To continue, we use $|\boldsymbol{a}\times \hat{\boldsymbol{n}}|^2=|\boldsymbol{a}|^2-(\boldsymbol{a}\cdot \hat{\boldsymbol{n}})^2$ for an arbitrary vector $\boldsymbol{a}$ and any unit vector $\hat{\boldsymbol{n}}$.  Moreover, by writing $|\Lambda\boldsymbol{r}|^2=\Tr{[\Lambda \boldsymbol{r}\boldsymbol{r}^\T \Lambda^\T]}$ and
$(\Lambda\boldsymbol{r}\cdot \hat{\boldsymbol{n}})^2=\hat{\boldsymbol{n}}^\T\Lambda\boldsymbol{r}\boldsymbol{r}^\T\Lambda^\T \hat{\boldsymbol{n}}$ and using $\int{(\boldsymbol{r}\boldsymbol{r}^\T) \dif^3r}=\frac{1}{5}\frac{4\pi}{3}\Id_{3}$, we get
\begin{eqnarray}\label{Q-l-one}
\mathcal{Q}_{C_{l_1}^2}(\mathcal{E})&=&\min_{\hat{\boldsymbol{n}}}\left(\Tr{[\mathcal{M}]}-\hat{\boldsymbol{n}}^\T \mathcal{M} \hat{\boldsymbol{n}}\right),
\end{eqnarray}
where $\mathcal{M}=\frac{1}{2}(\Lambda\Lambda^\T+5{\boldsymbol{t}}{\boldsymbol{t}}^\T)$. The minimum is achieved  when $\hat{\boldsymbol{n}}$ is an eigenvector of $\mathcal{M}$ corresponding to the largest eigenvalue, so that
\begin{eqnarray}\label{NSqubitfinal}
\mathcal{Q}_{C_{l_1}^2}(\mathcal{E})&=&\mathcal{M}_2+\mathcal{M}_3,
\end{eqnarray}
where $\mathcal{M}_1\ge \mathcal{M}_2 \ge \mathcal{M}_3$  are  eigenvalues of $\mathcal{M}$ in nonincreasing order. Equation  \eqref{NSqubitfinal} provides an explicit expression for the quantumness of qubit channels.

The quantumness defined by Eq. \eqref{NSqubitfinal}  is bounded above by $1$. To see this note  from Eq. \eqref{Q-l-one} that
\begin{eqnarray}\label{Q-UpperB1}
\mathcal{Q}_{C_{l_1}^2}(\mathcal{E})&\le&\left(\Tr{[\mathcal{M}]}-\hat{\boldsymbol{n}}^\T \mathcal{M} \hat{\boldsymbol{n}}\right),
\end{eqnarray}
holds for any  unit vector $\hat{\boldsymbol{n}}$. Recalling that the trace of a matrix is independent of the basis chosen, one can define $\{\hat{\boldsymbol{n}},\hat{\boldsymbol{n}}_{\perp},\hat{\boldsymbol{n}}^{\prime}_{\perp}\}$ as a set of orthonormal basis, so
$\mathcal{Q}_{C_{l_1}^2}(\mathcal{E})\le \left(\hat{\boldsymbol{n}}^\T_{\perp} \mathcal{M} \hat{\boldsymbol{n}}_{\perp}+\hat{\boldsymbol{n}}^{\prime\T}_{\perp} \mathcal{M} \hat{\boldsymbol{n}}^{\prime}_{\perp}\right)$. This  holds for any orthonormal basis, hance  for any basis with  $\hat{\boldsymbol{n}}_{\perp}\cdot {\boldsymbol{t}}=\hat{\boldsymbol{n}}^{\prime}_{\perp}\cdot {\boldsymbol{t}}=0$ we find
\begin{eqnarray}\label{Q-UpperB2}
\mathcal{Q}_{C_{l_1}^2}(\mathcal{E})&\le& \frac{1}{2}\left(\hat{\boldsymbol{n}}^\T_{\perp} \Lambda\Lambda^\T \hat{\boldsymbol{n}}_{\perp}+\hat{\boldsymbol{n}}^{\prime\T}_{\perp} \Lambda\Lambda^\T \hat{\boldsymbol{n}}^{\prime}_{\perp}\right)\le 1.
\end{eqnarray}
Here, the last inequality follows from the fact that singular values of $\Lambda$ cannot be  greater than 1, so the expectation value of $\Lambda\Lambda^\T$ on any normalized vector is bounded above by $1$.

Interestingly, the unitary channels are the only channels that reach the maximum quantumness $1$. Recalling that quantumness of the unitary channels is equal to the one of the identity channel, we have to prove the above assertion for the   identity channel.  Obviously, for the identity channel for which  $\Lambda=\Id_3$ and $\boldsymbol t=0$, the bound is saturated. On the other hand, if  $\mathcal{Q}_{C_{l_1}^2}(\mathcal{E})=1$ then $\hat{\boldsymbol{n}}^\T_{\perp} \Lambda\Lambda^\T \hat{\boldsymbol{n}}_{\perp}=\hat{\boldsymbol{n}}^{\prime\T}_{\perp} \Lambda\Lambda^\T \hat{\boldsymbol{n}}^{\prime}_{\perp}=1$, implies that the channel should be unital, i.e. $\boldsymbol{t}=0$,  as the singular values of $\Lambda$ cannot reach  its maximum value $1$ unless $\boldsymbol{t}=0$. Putting $\boldsymbol{t}=0$, there will be no condition on $\hat{\boldsymbol{n}}_{\perp}$ and $\hat{\boldsymbol{n}}^{\prime}_{\perp}$, as such  $\Lambda=\Id_3$ and the channel is identity.

In the rest of this paper,  we drop the subscript ${C_{l_1}^2}$ from the quantumness and denote the quantumness, given by Eq.  \eqref{Q-l-one},  by $\mathcal{Q}(\mathcal{E})$  for the sake of brevity.
Next,   we go into more details about  qubit channels and their local action on a two-qubit system.

\section{Local action of the quantum channels}\label{correlations}
Equation \eqref{NSqubitfinal} provides an analytical expression for the quantumness of qubit channels and should facilitate the investigation of questions concerning the quantum channels. In particular, it may be helpful to investigate questions concerning  local actions of quantum channels such as the following;   Will a channel with high nonclassicality  will  produce more quantum effects when the channel acts locally on  composite systems?   Is there any  relation between the channel quantumness and the quantum correlation preserved under local action of the channel?  To address these questions, we need to   recall some important classes of  quantum channels with special  focus  on their local action. A quantum channel $\mathcal{E}$ is called an entanglement-breaking channel,  i.e., channels that wash out entanglement when  applied locally \cite{entanglementbreaking},  if and only if it can be written in the so-called Holevo form \cite{holevo},
\begin{equation}\label{holevo}
\mathcal{E}(\rho)=\sum_n R_n\Tr(\rho F_n),
\end{equation}
where $R_n$ is a density matrix and $\{F_n\}$ forms a POVM.
An important subclass  of the entanglement-breaking channels is the so-called quantum-classical (QC) channels  \cite{holevo}, defined by Eq. \eqref{holevo}  when $R_n=\project n$, where $\{\ket n\}$ is an orthonormal basis.   As it is clear from the definition, any QC channel is semiclassical. On the other hand, any semiclassical channel is at least coherence-breaking, with respect to some  basis, but the latter is a subset of QC channels  \cite{coherencebreaking}. Accordingly, a quantum channel is quantum-classical if and only if it is semiclassical.

Considering the effect of  semiclassical channels in removing quantum aspects of input states, it has been mentioned that local semiclassical channels cannot create discord type correlations. It follows from  the equivalency between   quantum-classical and  semiclassical channels  that there is a much  stronger fact: semiclassical channels  not only cannot create quantum correlations locally but also destroy quantum correlations of the party they are applied on \cite{correlationbreaking}.    Even  more, only these channels have such an effect. In summary, the following equivalent statements will clarify  semiclassical channels \cite{entanglementbreaking,correlationbreaking}; (i) $\mathcal{E}$ is a semiclassical or QC channel, (ii) $\mathcal{E}$ is a discord-breaking channel.
(iii) $(\mathfrak{I}\otimes\mathcal{E})\ket\beta\bra\beta$ is a zero-discord state on part $B$ where $\ket\beta=d^{-1/2}\sum\ket{ii}$ is a maximally entangled state.

The above discussion suggests that for preserving quantum correlations in a composite system, the assumed quantum channel should retain some   quantumness in the whole space of the one-party system. In the following, we will show that there is actually  an interesting  relation between the channel quantumness  and the  quantum correlation preserved under the local action of the channel. Not surprisingly, such  relation is based on the Choi-Jamio\l kowski isomorphism,  stating that there is a one-to-one correspondence between  quantum channel $\mathcal{E}$ and quantum state  $\rho_{\mathcal{E}}=(\mathcal{I}\otimes\mathcal{E})\project\beta$   \cite{mathematical}.
\begin{observation}\label{Obs-GQinequality} The following relation exists between quantumness of $\mathcal{E}$, measured by Eq. \eqref{NSqubitfinal}, and quantum correlation of $\rho_{\mathcal{E}}$, measured by the (normalized) geometric discord
\begin{equation}\label{GQinequality}
D_G(\rho_{\mathcal{E}})\leq\mathcal{Q}(\mathcal{E})\leq1.
\end{equation}
\end{observation}
To prove this, we first need to assert the following lemma.
\begin{lemma}\label{Lemma-Bobeffect}
Let $\mathcal{E}$ be a qubit channel described by the affine parameters  $\Lambda$ and $\boldsymbol t$. Then $\mathcal{E}$ acts locally on  Bob's  side  of a two-qubit state $\rho_{AB}$ as
\begin{eqnarray}
\label{Bobeffect}
\nonumber(\mathcal{I}\otimes\mathcal{E})\rho_{AB}=\frac{1}{4}\big(&&\Id_2\otimes\Id_2+\Id_2\otimes(\Lambda\boldsymbol{y}+\boldsymbol{t}).\boldsymbol{\sigma}+\\ &&\boldsymbol{x}.\boldsymbol{\sigma}\otimes\Id_2+\sum_{ij}T_{ij}^b\sigma_i\otimes\sigma_j\big),
\end{eqnarray}
where  $T^b=T\Lambda^t+\boldsymbol x\boldsymbol t^t$. Above $\boldsymbol x$,   $\boldsymbol y$   are the Bloch vectors of Alice and Bob, respectively,   and $T$ is   the correlation matrix of the initial state.
\end{lemma}
See   Appendix \ref{Appendix2} for a proof of the above lemma.

Now, using the above lemma, Observation \ref{Obs-GQinequality} can be straightforwardly proven by noting the following: (i) For $\rho_{AB}=\project\beta$, with $\ket\beta=(\ket{00}+\ket{11})/\sqrt{2}$,  we have $\boldsymbol x=\boldsymbol y=0$ and $T=\diag\{1,-1,1\}$, so Eq. \eqref{Bobeffect} reduces to
\begin{eqnarray}\label{betatransformation}
\rho_{\mathcal{E}}&=&(\mathcal{I}\otimes\mathcal{E})\project\beta \\ \nonumber &=&\frac{1}{4}\big(\Id_2\otimes\Id_2+\Id_2\otimes\boldsymbol{t}.\boldsymbol{\sigma}+\sum_{ij}\Lambda_{ij}^{\prime t}\sigma_i\otimes\sigma_j\big),
\end{eqnarray}
where  $\Lambda^\prime$ equals  $\Lambda$ except for the sign of its second column which is opposite to the one of $\Lambda$.
(ii)   The normalized geometric discord \cite{vedral} of $\rho_{\mathcal{E}}$ is equal to the sum of two lower eigenvalues of $\mathcal{N}=(\Lambda^{\prime}\Lambda^{\prime t}+\boldsymbol t\boldsymbol t^t)/2=(\Lambda\Lambda^t+\boldsymbol t\boldsymbol t^t)/2$. Recalling  $\mathcal{M}=(\Lambda\Lambda^t+5\boldsymbol t\boldsymbol t^t)/2$ we get $\mathcal{M}=\mathcal{N}+2\boldsymbol t\boldsymbol t^t$. Evidently   $\mathcal{N}\le \mathcal{M}$,   \cite{HornBook1985} which completes the proof of the observation.

Now,  several properties of inequality \eqref{GQinequality} are in order. (i) Looking at two matrices $\mathcal{M}$ and $\mathcal{N}$ shows that  $D_G(\rho_{\mathcal{E}})=\mathcal{Q}(\mathcal{E})$ if and only if  $\boldsymbol{t}=0$ (unital channels) or $\boldsymbol{t}$ is an eigenvector of $\mathcal{N}$ corresponding to its largest eigenvalue.
(ii) $D_G(\rho_{\mathcal{E}})=0$ if and only if $\mathcal{Q}(\mathcal{E})=0$, which follows from the fact that both  quantities   vanish if and only if  $\textrm{rank}\{\mathcal{N}\}=\textrm{rank}\{\mathcal{M}\}=1$, that is to say,  if and only if   there exists a unit vector $\hat{\boldsymbol{n}}\in \mathbb{R}^3$ such that  $\hat{\boldsymbol{n}}\hat{\boldsymbol{n}}^\T \boldsymbol{t}=\boldsymbol{t}$ and $\hat{\boldsymbol{n}}\hat{\boldsymbol{n}}^\T \Lambda=\Lambda$.  This was an expected result since QC channels are the only class of channels removing quantum discord and transforming any state into a quantum-classical state.
(iii) Both quantities attain their maximum value $1$ if and only if $\Lambda$ is an orthogonal matrix (consequently   $\boldsymbol{t}=0$), i.e. for the  unitary channels which transform a maximally entangled state into a maximally entangled state. (iv) Both quantities are invariant under (local) unitary maps.

Paying attention to $\mathcal{Q}(\mathcal{E})$ and its common features with geometric discord we  argue that, for the states defined  by  Eq. \eqref{betatransformation}, this quantity has the properties  necessary for a good measure of quantum correlation. The duality between $\mathcal{E}$ and $\rho_{\mathcal{E}}$,  described by the Choi-Jamio\l kowski isomorphism, reveals  that the quantumness  of $\mathcal{E}$  is a necessary resource for quantum correlation of $\rho_{\mathcal{E}}$. In other words, the preserved quantum correlation, measured by geometric discord, cannot exceed  the quantumness of the channel. Moreover, the fact  $D_G(\rho_{\mathcal{E}})=0$ if and only if $\mathcal{Q}(\mathcal{E})=0$, states  that a channel with nonzero quantumness will preserve some quantum correlation.
This sheds  light on  the claim that the more nonsemiclassical a quantum channel is, the more quantum correlation is preserved when the channel acts locally on a composite system. A  semiclassical channel, on the other hand,  destroys all one-party quantumness, and as such it cannot create or preserve any discord-like quantum correlation.

\begin{figure}[t!]
\centering
\includegraphics[width=8.2cm]{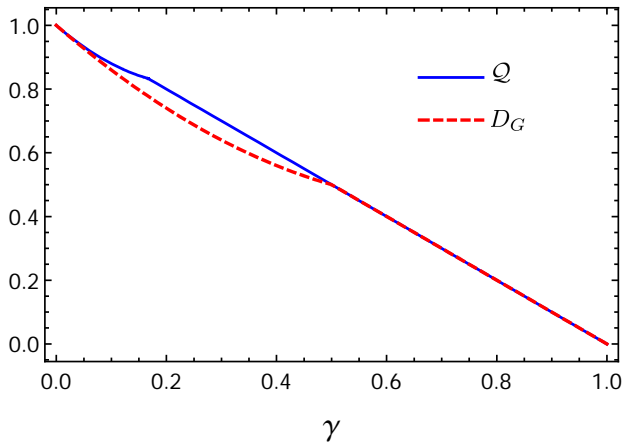}
\caption{(Color online) The  quantumness and geometric discord of the amplitude damping channel as a function of the channel's parameter $\gamma$.
For quantumness and geometric discord, the optimized $\hat{\boldsymbol{n}}$  shows the sudden change from $xy$-plane to $z$-axis on $\gamma=1/6$ and $\gamma=1/2$, respectively. }
\label{Fig-AmplitudeD}
\end{figure}

To demonstrate this, let us   consider   the amplitude damping channel characterized by the following Kraus operators \cite{nielsen}
\begin{equation}
K_1=
\begin{pmatrix}
    1       &0\\
    0       &\sqrt{1-\gamma} \\
\end{pmatrix},\quad K_2=
\begin{pmatrix}
    0       &\sqrt{\gamma}\\
    0       &0 \\
\end{pmatrix}.
\end{equation}
The affine parameters associated with this channel are  $\Lambda=\diag\{\sqrt{1-\gamma},\sqrt{1-\gamma},1-\gamma\}$ and $\boldsymbol{t}=(0,0,\gamma)^\T$.
Straightforward calculations show that for this channel the quantumness  and  geometric discord are given by
\begin{equation}
\mathcal{Q}(\mathcal{E}_a)=
\begin{cases}
\frac{1}{2}(6\gamma^2-3\gamma+2)&\gamma \leq\frac{1}{6}, \\
1-\gamma&\gamma>\frac{1}{6},
\end{cases}
\end{equation}
and
\begin{equation}
D_G(\rho_{\mathcal{E}_a})=
\begin{cases}
\frac{1}{2}(2\gamma^2-3\gamma+2)&\gamma \leq\frac{1}{2}, \\
1-\gamma&\gamma>\frac{1}{2},
\end{cases}
\end{equation}
 respectively.  Figure \ref{Fig-AmplitudeD} shows the quantumness  and geometric discord for this channel in terms of the channel's parameter $\gamma$. As can be seen from this figure, both  quantumness and geometric discord are  monotonically decreasing from  $1$ to $0$ as the channel's parameter $\gamma$ goes from $0$ to $1$. Indeed, when the damping rate $\gamma$ passes from $1/2$, quantum correlation reaches the channel's quantumness.

For entanglement-breaking channels,  although they  may preserve a fraction of one-party quantumness,  a significant amount of quantumness should be removed. More precisely, for  such channels the quantumness  is  bounded above as
\begin{eqnarray}\label{entbreaking}
\mathcal{Q}(\mathcal{E})&\leq & \frac{1}{2}\big(1-\hat{\boldsymbol{n}}^{\T}\Lambda\Lambda^\T\hat{\boldsymbol{n}}-\sum_{i\ne j}|\lambda_i\lambda_j|\big)\leq \frac{1}{2},
\end{eqnarray}
which holds for any unit vector $\hat{\boldsymbol{n}}$ which lies in the direction of $\boldsymbol{t}$. This follows easily from  Eq. \eqref{Q-UpperB2} and noting that for entanglement-breaking channels we have \cite{entanglementbreakingqubit} $\sum_i|\lambda_i|\leq 1$,  leads to $\Tr[\Lambda\Lambda^\T]=\sum_i|\lambda_i|^2\leq 1-\sum_{i\ne j}|\lambda_i\lambda_j|$. Clearly, both of the inequalities in Eq. \eqref{entbreaking} cannot be saturated simultaneously, so $\mathcal{Q}(\mathcal{E})$  does not  definitely  reach  $1/2$ for an entanglement-breaking channel.
On the other hand,  for the unital channels whose necessary and sufficient condition  to be entanglement-breaking is $\sum_i|\lambda_i|\leq 1$ \cite{entanglementbreakingqubit}, the quantumness is bounded above  by $1/8$, saturated for example by a channel described by the Kraus operators $K_1=\Id_2/\sqrt{2}$, $K_2=\sigma_1/2$ and $K_3=\sigma_2/2$.

Before we conclude this section,  let us mention here that the quantity defined above measures the amount of quantumness preserved in the state during the action of the channel.  Clearly,  a nonzero value for the channel's quantumness does not offer an ability to  create quantum correlation.
For example, although unital qubit channels mostly preserve quantumness of the input states, they cannot create quantum correlation locally \cite{streltsovprl2011}.
On the other hand, a  nonunital channel  described by the Kraus operators $K_1=\ket+\bra1$ and $K_2=\project0$ creates the  maximum amount of quantum discord locally \cite{correlating power}, although its quantumness is only $\mathcal{Q}(\mathcal{E})=1/4$.
In summary,  for the creation of quantum correlations by a local channel, other additional properties rather than preserving quantumness are required.


\section{Application to quantum teleportation}\label{Section-Application}
Consider the generalized depolarizing  channel described by the  Kraus operators $K_i=\sqrt{p_i}\sigma_i$ for $i=0,1,2,3$ where $\sigma_0=\Id_2$, and $\sigma_1$, $\sigma_2$ and $\sigma_3$ are the Pauli matrices. For this channel, the affine parameters  are given by
$\Lambda=\diag\{\lambda_1,\lambda_2,\lambda_3\}$ and $\boldsymbol{t}=0$, where $\lambda_1=p_0+p_1-p_2-p_3$, $\lambda_2=p_0-p_1+p_2-p_3$, and $\lambda_3=p_0-p_1-p_2+p_3$. If $p_0\geq p_1\geq p_2\geq p_3$ then $\lambda_1\geq\lambda_2\geq\lambda_3$ and we get for quantumness of this  channel  $\mathcal{Q}(\mathcal{E}_{\textrm{GD}})=\frac{1}{2}(\lambda_2^2+\lambda_3^2)=(p_0-p_1)^2+(p_2-p_3)^2$.

The generalized depolarizing channel is of particular importance because of  its usefulness in studying quantum teleportation. The   teleportation protocol \cite{BennettPRL1993} consists of a two-qubit states $\rho_{AB}$ shared between two separated parties, say Alice and Bob, and an unknown qubit state $\rho_{\textrm{in}}=\ket{\psi_{\textrm{in}}}\bra{\psi_{\textrm{in}}}$ where $\ket{\psi_{\textrm{in}}}=\cos{\vartheta/2}\ket{0}+\e^{i\varphi}\sin{\vartheta/2}\ket{1}$
with $0\leq\vartheta\leq\pi$ and $0\leq\varphi<2\pi$. The protocol is described by the generalized depolarizing channel
\begin{equation}\label{P0Protocol}
\mathcal{E}_{\textrm{GD}}(\rho_{\textrm{in}})=\sum_{i=0}^{3}p_{i}\sigma_{i}\rho_{\textrm{in}}\sigma_{i},
\end{equation}
where
$p_{i}=\Tr(E_i \rho_{AB})$ and
$\sum_{i}p_{i}=1$. Here $E_i=|\Psi_{\textrm{Bell}}^{i}\rangle\langle
\Psi_{\textrm{Bell}}^{i}|$ where $|\Psi_{\textrm{Bell}}^{i}\rangle$ are the four
maximally entangled Bell states associated with the Pauli matrices
$\sigma_i$, i.e. $E_i=(\sigma_i\otimes \sigma_0) E_0
(\sigma_i\otimes \sigma_0)$ for $i=0,\cdots,3$. Furthermore, for optimal utilization of a given entangled state as
resource, one has to  choose the local basis in such a way  that  $p_0=\max\{p_i\}$.

\begin{figure}[t!]
\centering
\includegraphics[width=8cm]{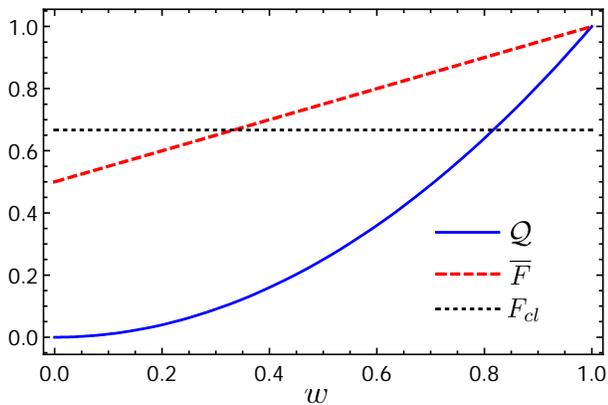}
\caption{(Color online) The  channel's quantumness and fidelity of the teleportation,   as a function of the  parameter $w$, when the Werner state is shared between Alice and Bob.  The horizontal line  shows the best classical fidelity $2/3$.  }
\label{Fig-Werner}
\end{figure}

To characterize the quality of the teleported state  $\mathcal{E}_{\textrm{GD}}(\rho_{\textrm{in}})$, it is useful to look at the fidelity
between $\rho_{\textrm{in}}$ and $\mathcal{E}_{\textrm{GD}}(\rho_{\textrm{in}})$. We find
$F(\rho_{\textrm{in}},\mathcal{E}_{\textrm{GD}}(\rho_{\textrm{in}}))=\frac{1}{2}(1+\hat{\boldsymbol{r}}^\T\Lambda \hat{\boldsymbol{r}})$ where
$\hat{\boldsymbol{r}}=\big(\sin{\vartheta}\cos{\varphi},\sin{\vartheta}\sin{\varphi},\cos{\vartheta}\big)^{\T}$ is the Bloch vector of the input qubit.
Moreover, the average fidelity is defined  by averaging the
fidelity  over all possible
input states ${\overline F}(\mathcal{E}_{\textrm{GD}})=\frac{1}{4\pi}\int_{0}^{2\pi}\textrm{d}\varphi
\int_{0}^{\pi}F(\rho_{\textrm{in}},\mathcal{E}_{\textrm{GD}}(\rho_{\textrm{in}}))\sin{\vartheta}\textrm{d}\vartheta$, and turns out that ${\overline F}(\mathcal{E}_{\textrm{GD}})=\frac{1}{3}(1+2p_0)$. It is clear that Alice and Bob could gain a fidelity better than $2/3$ (the  best possible fidelity when they  communicate only through classical channel), if and only if  $p_0>\frac{1}{2}$. A simple investigation shows that  if $p_0>\frac{1}{2}$ then $\mathcal{Q}(\mathcal{E}_{\textrm{GD}})\neq 0$. Equivalently, if $\mathcal{Q}(\mathcal{E}_{\textrm{GD}})=0$ then $p_0\leq\frac{1}{2}$ and we do not benefit from the quantum advantages. However, the inverse is not correct meaning that it is possible for the channel to possesses nonzero quantumness but the average fidelity of the teleportation to be less than $2/3$. This implies that in order to have a teleportation protocol with fidelity better than the classical one, a nonzero quantumness for the associated channel is necessary, although  it is not sufficient.  For example, for the  Werner state $\rho_{AB}=w\project\Psi+\frac{1-w}{4}\Id_4$, we have $\mathcal{Q}(\mathcal{E}_{\textrm{GD}})=w^2$ and ${\overline F}(\mathcal{E}_{\textrm{GD}})=\frac{1}{2}(1+w)$. This state is separable (disentangled) if and only if $w\leq 1/3$ and, not surprisingly, only for such values of $w$ for which the fidelity of the protocol is less than $2/3$. However, as it is clear from Fig. \ref{Fig-Werner}, even for  $w\leq 1/3$, the  channel possesses a nonzero quantumness.

\section{Conclusion}\label{conclusion}
Regarding that quantum coherence is the most fundamental aspect of quantum physics, it can be used to  characterize and quantify   other nonclassical features identified in the emerging  quantum information theory.   Using this very signature of quantumness, we propose the average coherence of  the channel's outputs,  minimized over all incoherent bases, as the  degree of nonclassicality of the channels, $\mathcal{Q}(\mathcal{E})$. This, intuitively, quantifies  the amount of quantumness preserved by a quantum channel and  measures the deviation of the assumed channel from the so-called semiclassical channels. Applying the square of $C_{l_1}(\rho)$ as a qualified measure of coherence for qubit systems,  an analytical expression is gained for any qubit channel. It is shown that $\mathcal{Q}_{C^2_{l_1}}(\mathcal{E})$ attains its maximum value $1$ for any unitary map, however, in other cases it cannot reach  its maximum value due to the contraction of the Bloch sphere to its associated channel's ellipsoid, hence decreasing occurs  in quantum properties.  By decrease in quantum properties we mean that the error of description of states  using a family of classical probability distributions is getting more and more close to being vanished when the measurement is in the minimized basis.
Obviously, doing maximization rather than minimization in our definition of quantumness introduces the basis that most  states are noncompatible with, meaning that  most  states show quantumness if we perform measurement in such a basis.  In the case of maximization, it could be straightforwardly proved that  the contraction into a totally mixed state has the lowest amount and  contraction into a pure state, which is a  semiclassical map, has the largest amount.

Moreover, utilizing the Choi-Jamio\l kowski isomorphism between the quantum channel $\mathcal{E}$ and Choi state $\rho_{\mathcal{E}}$, we show that the amount of quantum correlation preserved in the Choi state is bounded above  by the amount of quantumness of the quantum operation.   The bound is saturated both for the zero-discord states and the maximally entangled ones. Even more, the saturation occurs faithfully in a sense that one quantity  attains its  respective minimum (maximum)  if and only if the other one  attains its minimum (maximum).  This, in turn, proposes  to consider  the quantumness of $\mathcal{E}$ as  a necessary resource for the quantum correlation of $\rho_{\mathcal{E}}$ and suggests that such notion of quantumness can be considered as a quantum correlation measure for the class of Choi states.  For an entanglement-breaking channel whose local action washes  out all quantum entanglement of the input states, the preserved discord type quantum correlation of the Choi states cannot be larger  than the washed out one, meaning that   the channel should clear more than half the discord type correlations in order to remove all quantum  entanglement. Due to the widespread application of the quantum channels in  quantum information theory,  presenting a measure to quantify nonclassicality of the quantum channels could  be helpful in identifying the resources of the quantum advantages.  We  further introduce such application by applying our measure to quantum teleportation protocol and showing that the quantumness of the underlying channel is a necessary resource to gain fidelity better than the one offered by the best classical protocol.

Finally, it should be noted that averaging over pure states rather than all states in Eq. \eqref{NS} leads also to a measure quantifying quantumness (nonsemiclassicality) of the quantum channels. Quantumness obtained using such substitution in qubit systems through $C_{l_1}^2$ is equal to geometric discord of the Choi state gained by the assumed channel.

\section*{acknowledgment}
This work was supported by Ferdowsi University of Mashhad under Grant No.  3/45469 (1396/10/09).

\appendix
\section{Proof of the monotonicity of  $C_{l_1}^2(\rho)$}\label{Appendix1}
Here we are going to prove properties of the squared  $l_1$-norm of coherence qualifying it as a coherence measure in qubit systems.
A coherence measure $C(\rho)$  is a map from the set of states to the set of non-negative real numbers, with the following conditions imposed by   QRT \cite{pleniocoherence2014,Aberg2006,streltsovrev2017}. (a) \emph{Non-negativity}: $C(\rho)\geq0$ and $C(\rho)=0$ if and only if $\rho$ belongs to the set of incoherent states $\mathcal{I}$. (b) \emph{Monotonicity}: $C(\rho)\geq C(\sum_n\hat{K}_n\rho\hat{K}_n^\dagger)$ for any non-selective incoherent operation. (c) \emph{Strong monotonicity}: $C(\rho)\geq \sum_n p_nC(\rho_n)$ where $\rho_n=\hat{K}_n\rho\hat{K}_n^\dagger/p_n$ and $\hat{K}_n$ is an incoherent Kraus operator.   (d) \emph{Convexity}: $\sum_i p_iC(\rho_i)\geq C(\sum_i p_i\rho_i)$. Also it has been  proved that convexity and strong monotonicity together imply monotonicity \cite{pleniocoherence2014}.

Note that  the squared  $l_1$-norm of coherence  satisfies positivity and convexity as the square of any convex nonnegative function is convex too. It remains only to prove that  the aforementioned quantity  is strongly monotone under incoherent Kraus operators. In the qubit case, it has been shown  that a general incoherent operation admits a decomposition with at  most  $5$ Kraus operators  \cite{streltsov2017prl}
\begin{eqnarray}
K_1&=&
\begin{pmatrix}
    0       &b_1\\
    a_1       &0 \\
\end{pmatrix},\quad
 K_2=
\begin{pmatrix}
   a_2       &0\\
      0     &b_2\\
\end{pmatrix}, \\ \nonumber
 K_3&=&
\begin{pmatrix}
    a_3       &b_3\\
    0       &0 \\
\end{pmatrix}, \quad
K_4=
\begin{pmatrix}
    0       &0\\
    a_4       &b_4 \\
\end{pmatrix}, \quad
K_5=
\begin{pmatrix}
     a_5       &0\\
       0   &0 \\
\end{pmatrix},
\end{eqnarray}
where $a_i\in\mathbb{R}$, $b_i\in\mathbb{C}$, $\sum_{i=1}^5a_i^2=\sum_{i=1}^4|b_i|^2=1$, and $a_3b_3+a_4b_4=0$. For the input state $\rho=(\Id_2+\boldsymbol{r}\cdot \boldsymbol{\sigma})/2$, the squared $l_1$-norm coherence equals $C_{l_1}^2(\rho)=(r_1^2+r_2^2)/4$.  On the other hand
\begin{eqnarray}\label{SM1}
\sum_{i=1}^5p_iC_{l_1}^2\Big(\frac{K_i\rho K_i^\dagger}{p_i}\Big)&=&\sum_{i=1}^2p_iC_{l_1}^2\Big(\frac{K_i\rho K_i^\dagger}{p_i}\Big) \\ \nonumber
&=&w(q_i,q_i^\prime, \rho) C_{l_1}^2(\rho),
\end{eqnarray}
where
\begin{equation}
w(q_i,q_i^\prime, \rho)=\sum_{i=1}^2\frac{2q_iq_i^\prime}{q_i(1+r_3)+q_i^\prime(1-r_3)}
\end{equation}
and  we have defined  $q_i=a_i^2$ and  $q_i^\prime=|b_i|^2$ (for $i=1,2$) with  conditions  $q_1+q_2\le 1$ and $q_1^\prime+q_2^\prime\le 1$. In Eq. \eqref{SM1}, the first line follows from the fact that the three last  Kraus operators remove coherence of the input states. To complete the proof of monotonicity of $C_{l_1}^2(\rho)$, we need to show that  $w(q_i,q_i^\prime, \rho)$ cannot exceed 1.
A straightforward calculation can be applied to see that for the parameters $q_1=q_1^\prime=1-q_2=1-q_2^\prime$, the factor  $w(q_i,q_i^\prime,\rho)$ reaches its maximum value $1$. This completes the proof.


\section{Proof of the Lemma \ref{Lemma-Bobeffect}}\label{Appendix2}
To prove Eq. \eqref{Bobeffect}, we start with the Hilbert-Schmidt representation of a general two-qubit state
\begin{eqnarray}
\rho_{AB}=\frac{1}{4}\big(\Id_2\otimes\Id_2&+&\boldsymbol{x} \cdot\boldsymbol{\sigma}\otimes\Id_2 \\ \nonumber
&+& \Id_2\otimes\boldsymbol{y} \cdot\boldsymbol{\sigma}+\sum_{ij}T_{ij}\sigma_i\otimes\sigma_j\big).
\end{eqnarray}
Using this and the linear property of the quantum channels we get
\begin{eqnarray}\label{1}
\nonumber(\mathcal{I}\otimes\mathcal{E})\rho_{AB}=&&\frac{1}{4}\big(\Id_2\otimes\mathcal{E}(\Id_2)+\boldsymbol{x}.\boldsymbol{\sigma}\otimes\mathcal{E}(\Id_2)+\\ &&\Id_2\otimes\mathcal{E}(\boldsymbol{y}.\boldsymbol{\sigma})+\sum_{ij}T_{ij}\sigma_i\otimes\mathcal{E}(\sigma_j)\big).
\end{eqnarray}
In affine representation, one can easily shows that $\mathcal{E}(\Id_2)=\Id_2+\boldsymbol t.\boldsymbol\sigma$ and $\mathcal{E}(\sigma_i)=\sum_{j} \Lambda_{ji}\sigma_j$, where can be used in Eq. \eqref{1}  to get
\begin{eqnarray}
&&\nonumber(\mathcal{I}\otimes\mathcal{E})\rho_{AB}=\frac{1}{4}\big(\Id_2\otimes\Id_2+\Id_2\otimes(\Lambda\boldsymbol{y}+\boldsymbol{t}).\boldsymbol{\sigma}+\\ &&\boldsymbol{x}.\boldsymbol{\sigma}\otimes\Id_2+\boldsymbol{x}.
\boldsymbol{\sigma}\otimes\boldsymbol{t}.\boldsymbol{\sigma}+\sum_{il}(\sum_j\Lambda_{lj}T_{ij})\sigma_i\otimes\sigma_l\big).
\end{eqnarray}
After  rearranging the terms, we arrive at Eq. \eqref{Bobeffect}.


\end{document}